\documentclass[parskip=full,12pt]{article}

\usepackage{amssymb}
\usepackage{amsthm}
\usepackage{amsmath, bm}
\usepackage{mdwtab,booktabs}
\usepackage{graphicx}
\usepackage{natbib}
\usepackage{url}
\usepackage{color,soul}
\usepackage{multirow}
\usepackage[top=1in, bottom=1in, left=1in, right=1in]{geometry}
\usepackage[linesnumbered,ruled]{algorithm2e}
\usepackage{tikz}
\usetikzlibrary{er,positioning,arrows.meta,calc}
\usepackage[labelformat=simple]{subcaption}
\usepackage{fancyhdr}
\usepackage{fancyvrb}
\usepackage{float}
\usepackage[makeroom]{cancel}
\usepackage{enumitem}

\newcommand{\E}{\mathbb{E}}

\usetikzlibrary{er,positioning,arrows.meta,arrows,shapes,calc}

\tikzstyle{block} = [rectangle, draw, fill=white, 
    text width=7cm, text centered, rounded corners, minimum height=3em]

\tikzset{
     arrow/.style = { thick,  ->, >=Triangle},
}

\def\spacingset#1{\renewcommand{\baselinestretch}%
{#1}\small\normalsize} \spacingset{1}

\title{Flexible Modeling of Hurdle Conway-Maxwell-Poisson Distributions with Application to Mining Injuries}

\author{Shuang Yin \thanks{Department of Statistics, University of Connecticut, 215 Glenbrook Road, Storrs, CT,  06269-4120, USA. Email: \texttt{shuang.yin@uconn.edu}.} 
\and Dipak K. Dey \thanks{Department of Statistics, University of Connecticut, 215 Glenbrook Road, Storrs, CT,  06269-4120, USA. Email: \texttt{dipak.dey@uconn.edu}.}
\and Emiliano A. Valdez \thanks{Department of Mathematics, University of Connecticut, 341 Mansfield Road, Storrs, CT, 06269-1009, USA. Email: \texttt{emiliano.valdez@uconn.edu}.} 
\and Xiaomeng Li\thanks{Department of Statistics, University of Connecticut, 215 Glenbrook Road, Storrs, CT,  06269-4120, USA. Email: \texttt{xiaomeng.li@uconn.edu}.}}

\begin{document}
\date{}
\clearpage\maketitle
\thispagestyle{empty}

\begin{abstract}
While the hurdle Poisson regression is a popular class of models for count data with excessive zeros, the link function in the binary component may be unsuitable for highly imbalanced cases. Ordinary Poisson regression is unable to handle the presence of dispersion.  In this paper, we introduce Conway-Maxwell-Poisson (CMP) distribution and integrate use of flexible skewed Weibull link functions as better alternative. We take a fully Bayesian approach to draw inference from the underlying models to better explain skewness and quantify dispersion, with Deviance Information Criteria (DIC) used for model selection. For empirical investigation, we analyze mining injury data for period 2013-2016 from the U.S. Mine Safety and Health Administration (MSHA). The risk factors describing proportions of employee hours spent in each type of mining work are compositional data; the probabilistic principal components analysis (PPCA) is deployed to deal with such covariates. The hurdle CMP regression is additionally adjusted for exposure, measured by the total employee working hours, to make inference on rate of mining injuries; we tested its competitiveness against other models. This can be used as predictive model in the mining workplace to identify features that increase the risk of injuries so that prevention can be implemented.
\end{abstract}

\bigskip

\bigskip

\bigskip


\bigskip

\section{Introduction} \label{sec:intro}
\markboth{Introduction}{}

The mining industry poses some of the most dangerous workplace environment so that mining safety and health is always of utmost importance for mine operations management. Being able to accurately and appropriately measure the degree of riskiness of various mining functions can help mine workers decide where to work, and in the case of unionized members, to negotiate hazard-related salaries. It is not uncommon in practice to continually monitor the number of mine accidents and injuries, together with understanding the significant factors that drive these accidents and injuries.

There appears to be a number of research related to understanding and management of work injuries in the mining workplace. \citet{coleman2007measuring} examined the effectiveness of safety programs related to mining injuries with lost workdays. \citet{paul2009predictors} proposed a step-by-step multivariate logistic regression model to quantify the hazardousness and use this as a predictive model for the number of accidents. \cite{liu2019risk} described three management models aimed at stabilizing safety in the case of emergencies in the mining workplace. \citet{nowrouzi2017bibliometric} provides a bibliometric overview of the top 56 most cited articles on mining injuries.

The imbalanced distribution of majority (non-event) and minority (event) classes, which results in misleading predictions, poses a challenging task. Even though the information contained in the majority class is very important, the hazard rate is estimated and analyzed relying on the samples from the minority class. The consequences of overestimating or underestimating the hazard rate will directly impact the workers' safety and the mining company's financial well-being. Therefore, the study of the imbalanced problem is vital from the company's perspective.

Hurdle Poisson, proposed by \cite{mullahy1986hurdle}, is one of the most commonly used models to fit the count data that has an overly excess of zero counts. The traditional hurdle model has two parts: a binary model for estimating the excess zero counts and a Poisson model for the positive counts. The hurdle Poisson model has been widely deployed and investigated in several diverse fields, such as claims modeling for pricing in insurance (\cite{sarul2015genins}) and analyzing congressional responses to court decisions in sociology (\cite{zorn1998socio}), in which the excess zeros based on the hurdle Poisson models are more suitable than those of simple Poisson models. The binary model for the zero component uses either a logit or probit link, which is symmetric, to relate the response variable and the regressors. \cite{chen1999new} shows that asymmetric links may be more appropriate than symmetric links when the number of non-event is dramatically different from the number of event. With a special shape parameter, the asymmetric distribution such as skewed Weibull can estimate the skewness of the imbalanced data through incorporating the shape parameter. \cite{caron2018categorical} shows the flexibility of the skewed Weibull distribution when it is used as the link function in the generalized linear model for binary regression, and also shows the better outcomes compared with symmetric links. A related research by \cite{ma2006bayesian} overcame the weakness of inefficiency of the parameter estimates and lack of shared information by fitting a simple Poisson regression in a Bayesian framework; however, the underlying model often encounters the condition in which the accident events occur very rarely, which also leads to high skewness.

The ordinary Poisson regression model usually encounters problems caused by overdispersion, or sometimes, underdispersion. By incorporating a new parameter which controls the amount of dispersion, the Conway-Maxwell-Poisson (CMP) (\cite{shmueli2005useful}) can better evaluate the degree of dispersion of the data. The usual MCMC algorithm is unable to compute the accept ratio due to the normalized constant of CMP. Therefore, the exchange algorithm (\cite{chanialidis2018efficient}) is deployed, together with the density of auxiliary data sampled from the distribution estimated at the value of parameters from proposed distribution. With all assumptions satisfied, the posterior distribution for the parameters can be derived and based on which the Bayesian MCMC sampling will be performed.

Another statistical risk characteristics about the accident related factors, especially those pertaining to our mining data, are the proportions of employee hours for different types of mining work. This includes, for example, time spent for underground operations, surface operations, and office work. The total proportions for all types sum up to 100\%; the proportions are therefore analyzed in a routine of compositional data analysis. There has been some approaches to the solution of compositional vector problems, e.g., \cite{egozcue2003isometric} and \cite{barcelo2001mathematical}. However, most of these schemes can hardly accommodate zero values to reduce the dimensions. Based on a specific form of Gaussian latent variable, \cite{tipping1999probabilistic} formulated the Probabilistic Principal Component Analysis (PPCA), which can deal with the high sparsity in the covariates and is considered a generalization of the classical PCA. The classical PCA is a special case of PPCA when the covariance of the noise, which is derived from the distribution of the data after marginalizing out the latent variables, becomes infinitely small. The parameters of interest, weight matrix, and covariance of the noise, can be obtained by maximum likelihood estimation via the closed form derived by \cite{tipping1999probabilistic} and the MLE of the parameters are obtained by Expectation-Maximization (EM) algorithm (\cite{dempster1977maximum}). Because of its stability and fast convergence, the only stable local extreme value is the global maximum at which the true principal subspace is found. Another advantage of EM algorithm is that it is computationally efficient. A regularized version of the BIC is employed as a model selection tool to determine the optimal number of components in order to solve the issue of poor performance due to the occurrence of singularities for some starting values of the EM algorithm. This modified version of the BIC (\cite{fraley2007bayesian}) evaluates the likelihood at the maximum a posteriori (MAP) estimator instead of the MLE. The MAP is obtained in the M step where the likelihood function with a conjugate prior is maximized. 

In this paper, the parameter estimation of all the proposed models are taken in a fully Bayesian framework. The popular measure of model fit and comparison in a Bayesian perspective is tested by the Deviance Information Criterion (DIC) statistic. \cite{spiegelhalter2002bayesian} demonstrated that DIC includes how well the model fits the data (goodness of fit) and the complexity of the model (effective number of parameters). The Heidelberg and Welch Diagnostic is used to check that the Markov chain in the algorithm is from a stationary distribution. 

We have structured the rest of the paper as follows. In Section \ref{sec:ppca}, we describe the derivation and implementation of PPCA, a tool used to address the compositional nature of predictor variables in our mining data. In Section \ref{sec:hcrmodels}, we describe hurdle count regression models and show that it is a partition of two independent components: a binary model and a positive count data model. We briefly describe binary regression models and their associated link functions, emphasizing that skewed Weibull is much more suitable for handling imbalanced zero counts. We introduce the Conway-Maxwell-Poisson model, which generalizes the ordinary Poisson with an added parameter to handle dispersion. The simulation studies in Section \ref{sec:simu} display the recovery work and the flexibility of the proposed models. Section \ref{sec:real} provides for a detailed analysis of the real data set and is presented to show the performance of the proposed methodology. Section \ref{sec:conclude} provides conclusion. 

\section{Probabilistic principal components analysis} \label{sec:ppca}
\markboth{Probabilistic principal components analysis}{}

Unlike the classical PCA, which is derived by finding eigenvectors, the PPCA is calculated based on a probabilistic scheme. PPCA is indeed closely related to factor analysis (\cite{bartholomew2011latent}), except in factor analysis, the variance in the error term can be heterogeneous. It can be shown that classical PCA is a special case of PPCA.

Suppose $\{ \bm x_i\}_{i=1}^N$ are compositional vectors where $\bm x_i \in \mathcal{R}^d$ can be written as $\bm x_i = \bm B \bm z_i + \bm\mu + \bm\epsilon_i$,
where $\bm z_i \in \mathcal{R}^k$ is the latent variable, columns of $\bm B$ are the principal components, and $\bm B \in \mathcal{R}^d \times \mathcal{R}^k$, $\bm \mu \in \mathcal{R}^k$ and $\bm \epsilon_i$ is the error term with distribution $N(0, \sigma^2 \bm I)$. Thus the distribution of $\bm x_i$, given the latent variable $\bm z_i$, is
\begin{align*}
    \bm x_i | \bm z_i \sim N(\bm B \bm z_i + \bm \mu, \sigma^2 \bm I).
\end{align*}
Here we suppose $\bm z_i \sim N(\bm 0, \bm I)$ and $\bm x_i \sim N(\bm \mu, \bm B \bm B^T + \sigma^2 \bm I)$ with suitable dimensions.

The Bayes posterior distribution of $\bm z_i | \bm x_i$ is
\begin{align*}
    \bm z_i | \bm x_i \sim N(\bm B^T (\sigma^2 \bm I +  \bm B \bm B^T)^{-1}(  \bm x_i - \bm\mu), \bm I -\bm B^T (\bm B \bm B^T+\sigma^2 \bm I)^{-1}\bm B).
\end{align*}
The posterior variance of $\bm z_i$ is independent of $\bm x_i$ and we can derive PCA as follows. Start with
\begin{equation*}
    \lim_{\sigma^2 \rightarrow 0} \big[\bm I -\bm B^T (\bm B \bm B^T+\sigma^2 \bm I)^{-1}\bm B\big] \rightarrow 0, 
\end{equation*}
so that the posterior variance now is zero and  the projection matrix $\lim_{\sigma^2 \rightarrow 0} \big[(\bm B \bm B^T+\sigma^2 \bm I)^{-1}\bm B\big] = \bm B^T( \bm B \bm B^T)^{-1}$ is orthogonal. We can obtain the standard principal components as
$$\bm B^T(\bm B \bm B^T)^{-1}(\bm x_i -\mu).$$
Consequently, the log-likelihood function can be written explicitly as
\begin{align*}
    \ell(\bm x_1, \ldots, \bm x_N) = -\frac{N}{2}\log\{(2\pi)^d |\bm B \bm B^T + \sigma^2 \bm I|\} - \frac{1}{2} \sum_{i=1}^N (\bm x_i - \bm \mu) (\bm B \bm B^T + \sigma^2 \bm I)^{-1} (\bm x_i - \bm \mu)^T.
\end{align*}
We use the average value of $\{\bm x_i\}_{i=1}^N$ as the estimate of $\bm \mu$. Based on the log-likelihood function, the maximum likelihood estimator (MLE) of $\bm B$ and $\sigma^2$ can be derived explicitly. However, the result of MLE could be a saddle point unless the calculated eigenvectors represent the principal subspaces. Moreover, the computation of calculating MLE can be burdensome for high-dimensional datasets.

Alternatively, the value of $\bm B$ and $\sigma^2$ can be estimated by EM algorithm (\cite{mclachlan2008}). The EM algorithm can provide the maximum likelihood estimates of parameters by iteratively alternating E (expectation) and M (maximization) steps. The EM algorithm is widely used in the imputation of missing values, while the latent variable $\bm z_i$ can be considered as ``missing'' data.

\begin{figure}[H]
    \centering
    \includegraphics[width=0.58\linewidth]{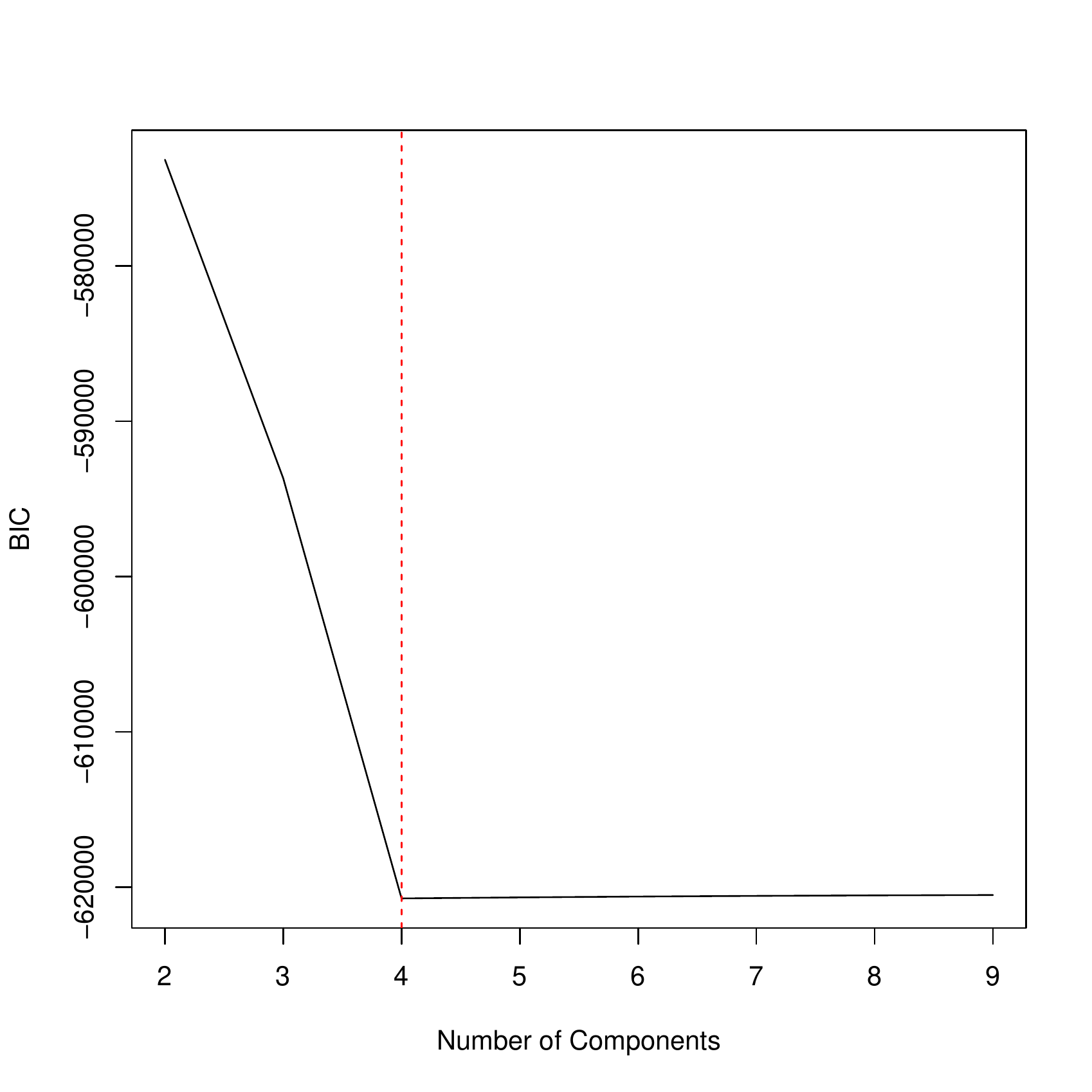}
    \caption{The selection of the number of principal components in PPCA via the BIC statistic}
      \label{fig:pca_bic}
\end{figure}

The Bayesian Information Criterion (BIC) is used to select the optimal number of principal components, and BIC is formulated as 
\begin{equation}
      \text{BIC} = -2\times \log \text{likelihood} + K \times \log m
\end{equation}
where $K$ is the number of free parameters in the model and $m$ is the sample size. The BIC can be interpreted as a criteria for model fit but with a penalty for model complexity. Based on this definition, the model which yields a smaller BIC is considered preferable. However, it is a noteworthy precaution that the BIC sometimes encounters the problem of noisy behavior resulting in performing poorly in model selection. The problem due to the EM algorithm with the singularities of randomly chosen starting points can be fixed by using modified version of BIC in which the log likelihood function evaluated at the maximum a posteriori (MAP) instead of the MLE. For a given number of components, the convolution of the likelihood and a conjugate prior are maximized at the M step (\cite{nyamundanda2010probabilistic}). Given the data drawn from MSHA discussed in Section \ref{sec:real}, Figure \ref{fig:pca_bic} shows that the BIC is minimized at four principal components. 

\section{Hurdle count regression models} \label{sec:hcrmodels}
\markboth{Hurdle count regression models}{}

Our dataset will be described as $D = (\boldsymbol{y},\boldsymbol{x},m)$ where $\boldsymbol{y}=(y_1,y_2,\ldots,y_m)'$ is a vector of responses, $\boldsymbol{x}=(\boldsymbol{x}_1,\boldsymbol{x}_2,\ldots,\boldsymbol{x}_m)'$ is a matrix of predictor variables, and $m$ is the total number of observations. The response variable $y$ is a count random variable. We define the positive component of $y$ as binary outcome of $y^+ = I(y>0)$ and denote the observations of positive counts and zero counts as $\boldsymbol{y}^+=(y^+_1,y^+_2,\ldots,y^+_m)'$ consisting of 1's and 0's.

Suppose that $N$ is a count random variable with probability mass function $P_N(n| \gamma, \boldsymbol{x})$ for $n=0,1,\ldots $ where $\gamma$ is a vector of parameter coefficients corresponding to the predictor variables. A hurdle count regression model for $y$ can be viewed as a two-partition model with the first partition for the zero component as a binary outcome and the second partition as the positive count data component. Its probability mass function can be expressed as
\begin{equation} \label{eq:hurdle}
P(y =n) = \begin{cases}
 1-p, & \text{if } n=0 \\
 p P^*_N(n|\bm \gamma, \boldsymbol{x}), & \text{if } n=1,2,\ldots \\
 \end{cases}
 \end{equation}
where $\bm \gamma$ is a vector of parameters $P^*_N(n|\bm \gamma, \boldsymbol{x}) = P_N(n|\bm \gamma, \boldsymbol{x})/\sum_{k=1}^{\infty} P_N(n|\bm \gamma, \boldsymbol{x})$ is the probability mass function of the modified positive count random variable. It can easily be deduced that $p = P(y>0) = P(y^+=1)$, so that its complement is $1-p = P(y=0) = P(y^+=0)$. Note that $p$ may also depend on the set of predictor variables $\boldsymbol{x}$ with a set of coefficients $\boldsymbol{\beta}$.

The hurdle model is one approach to handle the excessive zeros in count data. It is based on the premise that the count data consists of zeros, the point of truncation, and the positive component. For the positive component, the count data is modeled with a zero-truncated count distribution. Another widely popular approach that may be suitable for excessive zeros is the use of zero-inflated models. See \cite{lambert1992zero}. In contrast to hurdle models, zero-inflated models are viewed as a mixture distributions of zeros and positive count data components. The two models are equivalent, with the primary difference only on the parameterization. Within the scope of our application, the hurdle model is much more intuitively explainable and interpreted.

\subsection{Link functions for the binary component} \label{sec:binary}

The binary component of the hurdle model will be described as a binary regression model based on its latent variable interpretation. The binary regression model falls within the class of generalized linear models (GLMs) with link function $g(\cdot)$ that relates the probability of event to the set of predictor variables as $g(p_i) = \boldsymbol{x}'_i \boldsymbol{\beta}$ where $\boldsymbol{\beta}=(\beta_0,\beta_1,\beta_2,\ldots,\beta_p)'$ is a vector of coefficients. In effect, we can conveniently express $p_i = g^{-1}(\boldsymbol{x}'_i \boldsymbol{\beta}) = p(\boldsymbol{x}'_i \boldsymbol{\beta})$. It is quite common to include an intercept coefficient $\beta_0$ in the linear relationship so that without loss of generality, we can assume that the vector of predictor variables is augmented as $(1,\boldsymbol{x}_{i1},\ldots,\boldsymbol{x}_{ip})'$. See \cite{mccullagh1989} and \cite{cameron2013}.

According to the latent variable interpretation, $y^+_i$, for observation $i$, is related to an unobserved variable $z_i$ as $y^+_i = I(z_i>0)$. $z_i$, also called the latent variable, is directly linked to the predictor variables as a linear model with an error component as $z_i = \boldsymbol{x}_i \boldsymbol{\beta} + u_i$, where $\boldsymbol{\beta}$ is a vector of coefficients for the predictors and the error component $u_i|\boldsymbol{x}'_i \sim F$. Here $F$ is the distribution function of $u_i$, given $\boldsymbol{x}'_i$, and it therefore follows that
\begin{equation}
\begin{split}
p(\boldsymbol{x}'_i \boldsymbol{\beta})  &= \text{Prob}(y_i^+=1|\boldsymbol{x}_i) = \text{Prob}(z_i >0)  \\
        &= \text{Prob}(u_i > - \boldsymbol{x}'_i \boldsymbol{\beta}) = 1-F(-\boldsymbol{x}'_i \boldsymbol{\beta}). \label{eq:latent}
\end{split}
\end{equation}
When $F$ is the distribution function of a symmetric random variable $u_i$ with mean 0, we have $p(\boldsymbol{x}'_i \boldsymbol{\beta}) = F(\boldsymbol{x}'_i \boldsymbol{\beta})$. In this case, $F^{-1}$ determines the link function in the GLM framework.

For our purposes, we consider the following families of link functions (without confusion, the subscript $i$ has been dropped):
\begin{itemize}
\item[(1)] \textit{Probit regression}: When $u$ has a normal distribution with mean 0 and scale parameter 1, i.e., $u \sim \text{Normal}(0,1)$, we have $F(u) = \Phi(u)$, where $\Phi$ is the distribution function of a standard normal. The link function is expressed as the probit function $g(p) = \Phi^{-1}(p)$.  This class of function belongs to symmetric link, and the resulting model is referred to as probit regression. Note that in the more general case where we have $u$ to be normal with mean 0 and variance $\sigma^2$, equation (\ref{eq:latent}) becomes
\begin{equation*}
p(\boldsymbol{x}'_i \boldsymbol{\beta})  = \Phi(-\boldsymbol{x}'_i \boldsymbol{\beta}/\sigma).
\end{equation*}
In this case, the coefficients are said to be scaled coefficients.
\item[(2)] \textit{skewed Weibull link function}: The distribution function of skewed Weibull has the form 
\begin{equation}
F_{\text{SW}}(u) = 1- \exp[-(u/\sigma)^\alpha] \label{eq:SW}
\end{equation}
defined only for $u>0$, provided $\sigma>0$ and $\alpha >0$; its value will be zero elsewhere. When $\sigma=1$, this becomes the standard form of the Weibull distribution and the corresponding link function can be expressed as $g(p) = [-\log(1-p)]^{1/\alpha}$. This class of function belongs to asymmetric link, which has been considered to work well for imbalanced binary outcomes.
\end{itemize}

For other link functions, see \cite{Yin2020skewed}. It is noteworthy to mention that when comparing the two link functions above, we need to set equivalent variances so that units of the coefficients are comparable. See \cite{long1997}. For example, if $\sigma =1$ in the probit link function, the corresponding coefficients in the skewed Weibull should be adjusted with the scale parameter
\begin{equation}
\sigma = 1/\sqrt{\Gamma(1+(2/\alpha)) - (\Gamma(1+(1/\alpha))^2}. \label{eq:scaleSW}
\end{equation}
Similarly, if we set $\sigma=1$ in the skewed Weibull link function, the corresponding coefficients in the probit link should be adjusted with the scale parameter
\begin{equation}
\sigma = \sqrt{\Gamma(1+(2/\alpha)) - (\Gamma(1+(1/\alpha)))^2}. \label{eq:scprobit}
\end{equation}

\subsection{Conway-Maxwell-Poisson distributions} \label{cmp}

The count random variable $N$ follows a Conway-Maxwell-Poisson (CMP) distribution if its probability mass function can be expressed as
\begin{equation}\label{equ:compois}
P_N(n|\lambda,\nu) = \frac{1}{Z(\lambda, \nu)} \frac{\lambda^n}{(n!)^{\nu}}, \ \text{for } n=0,1,\ldots,
\end{equation}
where
\[
Z(\lambda, \nu) = \sum_{j=0}^{\infty} \frac{\lambda^j}{(j!)^{\nu}},
\]
and $\nu \ge 0$ is a parameter. The normalizing constant $Z(\lambda,\nu)$ does not have an explicit expression, and may have to be numerically evaluated. The $\lambda$ parameter should be $\lambda>0$ when $\nu>0$, and $0 < \lambda < 1$ when $\nu=0$. We write $N \sim \text{CMP}(\lambda, \nu)$. It is straightforward to see that when $\nu =1$, we have the ordinary (standard) Poisson distribution with $P_N(n|\lambda) = e^{-\lambda} \lambda^n/n!$. The CMP distribution can be attributed to the work of \cite{conway1962CMP}.

While there are no explicit forms for the mean and variance of the CMP distribution, one can use the moment generating function, which can be verified to be equal to $\E(e^{Nt}) = Z(\lambda e^t, \nu)/Z(\lambda, \nu)$, to evaluate them. The parameter $\nu$ governs the level of dispersion in the CMP distribution. Recall that for the Poisson distribution, $P_N(n-1|\lambda)/P_N(n|\lambda)=n/\lambda$. For the CMP distribution, it can be shown that $P_N(n-1|\lambda,\nu)/P_N(n|\lambda,\nu)=n^{\nu}/\lambda$. Again we see that $\nu=1$, the CMP distribution becomes the ordinary Poisson distribution, which describes no dispersion. When $\nu < 1$, the rate of decay decreases less than Poisson and has a longer tail; this is the case of overdispersion. When $\nu >  1$, the rate of decay increases more in a nonlinear function, thus shortening the tail of the distribution; this is the case of underdispersion. For further details, please see  \cite{shmueli2005useful} and \cite{li2020}.

\subsubsection{Exchange algorithm}

All model parameters described in this paper used the Bayesian estimation method with approximation of the posterior distribution based on the popular Metropolis-Hastings algorithm, which is a type of a Markov Chain Monte Carlo (MCMC) method. The exception is the case of the standard CMP because the normalizing constant has to be estimated where we use the so-called exchange algorithm. For observation $i$, we incorporate the $p\times 1$ vector of predictors $\boldsymbol{x}_i=(1, x_{i1}, \dots, x_{ip})'$ through $\lambda_i$,  using the log link functions, $\log(\lambda_i) = \boldsymbol{x}_i'\boldsymbol{\beta}$, where $\boldsymbol{\beta}=(\beta_0, \beta_1, \dots, \beta_p)$ are the corresponding coefficients. In this case, we can rewrite the probability mass function of $N_i$ as $P_{N_i}(n|\mu_i)=h(n| \mu_i)/ Z_h (\mu_i)$ where $\mu_i=(\beta_i, \nu)$ and $h(n|\mu_i) = \lambda_i^{n}/n^{\nu}$.

The Bayesian inference for parameters $\boldsymbol{\beta}$ and $\nu$ in CMP regression models is a doubly-intractable problem, so the direct application of a standard MCMC method is infeasible. For example, a Metropolis-Hastings algorithm requires the calculation of the intractable ratios $\left \{Z_h(\mu_i)/Z_h(\mu_i^*)\right \}_{i=1}^m$ if it proposes a move from $\mu_i$ to $\mu_i^*$. Now the acceptance ratio becomes
\begin{equation*}
 \rho (\mu, \mu^*) = \min \left\{ 1, \frac{\prod_{i=1}^m \frac{h(n|\mu_i^*)}{Z_h(\mu_i^*)}}{\prod_{i=1}^m \frac{h(n|\mu_i)}{Z_h(\mu_i)}} \frac{q(\mu^*, \mu) \pi(\mu^*)}{q(\mu, \mu^*) \pi(\mu)}\right \}.
\end{equation*} 
Although \cite{shmueli2004modeling} proposed approximations by a truncated sum $Z_h(\mu_i)=\sum_{i=1}^k q_h(n|\mu_i)$ to estimate the normalized constant, there still exists some bias in the acceptance ratio. 

The acceptance ratio for the augmented posterior is now calculated as
\begin{align*}
    \rho_{exchange}(\mu, \mu^*)  &= \min \left\{ 1, \frac{\prod_{i=1}^m \frac{h(n_i|\mu_i^*)}{Z_h(\mu_i^*)}}{\prod_{i=1}^m \frac{h(n_i|\mu_i)}{Z_h(\mu_i)}} \frac{q(\mu^*, \mu) \pi(\mu^*)}{q(\mu, \mu^*) \pi(\mu)} 
    \frac{\prod_{n=1}^m \frac{h(n_i^* |\mu_i)}{Z_h (\mu_i)}}{\prod_{n=1}^m \frac{h(n_i^* |\mu_i')}{Z_h (\mu_i')}}\right \} \\
    & = \min \left\{1, \frac{\prod_{i=1}^m h(n_i|\mu_i^*)q(\mu^*, \mu)\pi(\mu^*)\prod_{i=1}^m h(n_i^* |\mu_i)}{\prod_{i=1}^m h(n_i|\mu_i)q(\mu, \mu^*)\pi(\mu)\prod_{i=1}^m h(n_i^* |\mu_i^*)} \xcancel{\frac{\prod_i \frac{1}{Z_h(\mu_i^*)}\prod_i \frac{1}{Z_h(\mu_i)}}{\prod_i \frac{1}{Z_h(\mu_i)}\prod_i \frac{1}{Z_h(\mu_i^*)}}} \right \}.
\end{align*}
The cancellation of the normalizing constants in the acceptance ratio above is due to the exchange of parameters $(\mu_i, \mu_i^*)$ associated with the data $\mathbf{N}=(n_1, ..., n_n)$ and the auxiliary data $\mathbf{N}^*=(n_1^*, ..., n_n^*)$, the auxiliary being discarded after each move. The acceptance ratio for the exchange algorithm becomes
\begin{align*}
    \rho (\mu, \mu^*)=\min \left \{1, \frac{\left \{ \prod_{i=1}^m h_{\mu^*}(n_i)  \right\} \pi (\mu^*) \pi(\gamma^*) \left \{ \prod_{i=1}^m h_{\mu}(n_i^*)  \right\}}{\left \{ \prod_{i=1}^m h_{\mu}(n_i) \right\}  \pi(\mu) \pi(\gamma)  \left \{ \prod_{i=1}^m h_{\mu^*}(n_i^*)  \right\}} \right \}.
\end{align*}

\subsubsection{Zero-truncated CMP for the positive count}

For the positive component of the hurdle model, we can easily show that the zero-truncated CMP distribution has the form:
\begin{equation}
P_N^*(n|\lambda,\nu) =  \frac{\frac{1}{Z(\lambda, \nu)}}{1-\frac{1}{Z(\lambda, \nu)}} \, \frac{\lambda^n}{(n!)^{\nu}} = \frac{1}{Z(\lambda, \nu)-1} \, \frac{\lambda^n}{(n!)^{\nu}}, \ \ \text{for } n=1,2,\ldots
\end{equation}
It is easy to see that in the special of the ordinary Poisson where $\nu=1$, we have the zero-truncated Poisson distribution with $P_N^*(n|\lambda,\nu) = [1/(e^{-\lambda}-1)] \lambda^n/n!$.

To incorporate predictors $\boldsymbol{x}$, we use the log link functions $\log (\lambda_i)= \boldsymbol{x}' \boldsymbol{\gamma}$. To generate posterior samples of the coefficients $\bm \gamma$ and $\nu$ in this zero-truncated model, we propose the candidate distributions $\boldsymbol{\gamma} \sim N(\boldsymbol{\gamma}^{t-1}, \boldsymbol{\Sigma})$ and $\nu \sim \log N(\nu^{t-1}, \sigma^2_{\nu})$ with acceptance ratio 
\begin{equation}
\begin{aligned}
    \rho_{\mu} = \min\left \{1, 
     \frac{\prod_{j:y_j>0} P_N^*(y_j|\mu^c)\pi(\mu^c)q(\mu^{t-1}|\mu^c)}{\prod_{j:y_j>0} P_N^*(y_j|\mu^{t-1})\pi(\mu^{t-1})q(\mu^c |\mu^{t-1})}
     \right\},
\end{aligned}
\end{equation}
where $\mu =(\gamma, \nu)$, $\mu^c$ is the candidate sample generated from the proposal distribution, and $\mu^{t-1}$ is the accepted samples from the $t-1$ step. As pointed out earlier, the normalized constant $Z$ of a CMP distribution does not have a closed form, conjugate priors are not available. The prior distribution of the dispersion parameter $\nu$ is chosen to be a lognormal distribution with a median at 1.

To be able to evaluate the zero-truncated probability $P_N^*(n|\lambda,\nu)$, we need to approximate the normalized constant term $Z(\lambda,\nu)$. The series $\lambda^j/(j!)^{\nu}$ converges for any $\lambda, \nu >0$ since $\lim_{j\rightarrow \infty} \lambda/j^{\nu}\rightarrow 0$ and $\lim_{j\rightarrow \infty} \lambda^j/(j!)^{\nu} \rightarrow 0$. Then there exists a value $K$ such that for $k > K$, $\lambda/k^{\nu} < 1$. Obviously, the infinite summation $Z(\lambda, \nu)$ can be written as $Z(\lambda, \nu) = \sum_{j=0}^k \frac{\lambda^j}{(j!)^{\nu}} + R_k$. Since $\lambda^j/(j!)^{\nu}$ decreases at a rate faster than a geometric series, then there exists a positive number $\epsilon_k \in (0, 1)$ such that for all $j >k$, $\frac{\lambda}{(j+1)^{\nu}}<\epsilon_k$ and $R_k=\sum_{j=k+1}^{\infty} \frac{\lambda^j}{(j!)^{\nu}}$ is bounded by $\frac{\lambda^{k+1}}{[(k+1)!]^{\nu}(1-\epsilon_k)}$ (\cite{minka2003computing}). Therefore by precautiously truncating the infinite summation and bounding the error, we can appropriately approximate $Z(\lambda,\nu)$.  

\subsection{Posterior distributions}

We need to evaluate the posterior distributions for Bayesian estimation. Recall that the hurdle model can be partitioned into independent zero component and positive count data component. The response variable for the zero part is a binary outcome and for positive count component, we use the zero-truncated count data distribution, which excluded zeros. Given our dataset $D = (\boldsymbol{y},\boldsymbol{x},m)$ and from (\ref{eq:hurdle}), we can run these two models in parallel because their posterior distributions are independent as shown below:
\begin{equation}
\begin{aligned}
\pi(\bm \beta, \bm \gamma|\bm y) 
& \propto  \prod_{i}^m [p_i \times P^*_N(y_i|\bm \gamma, \boldsymbol{x})]^{y_i^+} \times (1-p_i)^{1-y_i^+} \pi(\bm \beta, \bm \gamma) \\
&=  \prod_{i}^m p_i^{y_i^+} (1-p_i)^{1-y_i^+} \times [P^*_N(y_i|\bm \gamma, \boldsymbol{x})]^{y_i^+} \pi(\bm \beta) \pi(\bm \gamma)  \\
&=  \prod_{i}^m (1-F(-\bm x_i' \bm \beta))^{y_i^+} F(-\bm x_i' \bm \beta)^{1-y_i^+} \pi(\bm \beta) \times [P^*_N(y_i|\bm \gamma, \boldsymbol{x})]^{y_i^+} \pi(\bm \gamma)   \\
&= \pi(\bm \beta|\bm y^{+}) \times \pi(\bm \gamma|\bm y_{\text{zt}}).
\end{aligned}
\end{equation}
where $\boldsymbol{y}^+$ consists of 1's and 0's for the zero component and $\boldsymbol{y}_{\text{zt}} \subset \boldsymbol{y}$ is the zero-truncated subset of the entire dataset. The function $F(\cdot)$ is the distribution function associated with the link function of the binary model that relates the response and predictor variables $\bm x$ with coefficients $\bm \gamma$. This notation should not preclude the possibility of an additional parameter in the link function, which is true in the case of the skewed link function.

For our purposes, we assume the independent multivariate normal priors for $\bm \beta$ and $\bm \gamma$, respectively, with $\bm \beta\sim \text{N}_{p+1} (\boldsymbol{0}, \sigma^2_{\beta} \boldsymbol{I}_{p+1})$ and $\bm \gamma \sim \text{N}_{p+1} (\boldsymbol{0}, \sigma^2_{\gamma} \boldsymbol{I}_{p+1})$. The same set of predictor variables are used but the parameter values are different for the zero component and the zero-truncated. The probit link for the binary component is a symmetric link function with no extra parameter, but for the skewed Weibull link, there is the additional parameter $\alpha$ for which we will assume $\alpha \sim \text{Gamma}(0.1, 0.1)$ and independent of all other parameters.

\section{Simulation studies}\label{sec:simu}
\markboth{Simulation studies}{}

As hurdle models can be partitioned into independent binary and positive count data regressions, the primary goal of our simulation studies is to compare the different link functions of the binary regression and compare count data models fit on the simulated data sets generated from various models, including the probit and Weibull links, Poisson, and CMP regressions. The parameters incorporated in all the underlying models are fully estimated in a Bayesian analysis and the goodness of fit is measured by Deviance Information Criterion (DIC).

For model fitting in a Bayesian framework, we use the DIC which was proposed by \cite{spiegelhalter2002bayesian} to evaluate, compare, and assess the performance and quality of the models. The model deviance is defined as $-2 \times \log(f(\boldsymbol{y}|\theta))$, with $\theta$ to denote the vector of parameters, then
\begin{equation*}
\text{DIC} =2 \times \overline{\text{Deviance}}(\boldsymbol{y}, \theta) - \text{Deviance}(\boldsymbol{y}, \overline{\theta}),
\end{equation*}
where $\text{Deviance}(\boldsymbol{y}, \overline{\theta})$ is the deviance evaluated under the value of the posterior mean $\overline{\theta}$ of the corresponding parameters, $\overline{\text{Deviance}}(\boldsymbol{y}, \theta) = \frac{1}{n} \sum_{i=1}^n \text{Deviance}(\boldsymbol{y}, \theta^{(i)})$, which is the average estimated discrepancy for $N$ samples, and $\theta^{(i)}$ is the $i$th sample generated from the posterior distribution $\pi(\theta|D)$. We also use $p_D=\overline{\text{Deviance}}(\boldsymbol{y}, \theta) - \text{Deviance}(\boldsymbol{y}, \overline{\theta})$ to measure the effective model size and penalize for model complexity. The DIC is a Bayesian alternative to AIC and BIC. The model with smaller DIC generally exhibits better quality of fit to the data.

The simulation experiments have been generated according to the following procedure. First, we independently generate 1000 normally distributed random variables $x_{i1} \sim N(0, 1)$ for $i=1, 2, \dots, 1000$ and $\bm x_i =(1, x_{i1})'$. Then we independently generate four data sets each with 1000 observations separately from:
\begin{itemize}
\item[(1)] Bernoulli under probit link, with $\sigma=1$, relating the regression component according to $\bm x_i '\bm \beta = -1+ (-0.5) \times x_{i1}$;
\item[(2)] Bernoulli under Weibull link with $\sigma=1$, $\alpha=3$, and $\bm x_i '\bm \beta = -2 + 1 \times x_{i1}$;
\item[(3)] Poisson under log link with $\bm x_i '\bm \gamma = 1+ 0.3 \times x_{i1}$; and
\item[(4)] CMP with $\nu=0.63$ and under log link with $\bm x_i '\bm \gamma = 1+ 0.3 \times x_{i1}$.
\end{itemize}
For simulation 1, the simulated dataset is the element-wise multiplication of variables (1) and (3), while the for simulation 2, the data set is generated in a similar manner using variables (2) and (4). Both simulated datasets feature highly imbalanced distribution and the second dataset, generated from Weibull and CMP, has longer tail than the first dataset.

To summarize the posterior marginal densities of the parameters under a Bayesian framework, we use the highest posterior density interval which has the shortest length for a given probability content. A $100(1-\alpha)\%$ HPD region (\cite{chen1999monte}) for $\beta$ can be written as $\mathcal{C} =\{\beta: \pi(\beta | \text{data})\}$,
where $k$ is the largest number such that
\begin{align*}
\int_{\beta:\pi(\beta | \text{data}) \geq k} \pi(\beta|\text{data})d \beta = 1-\alpha,
\end{align*}
and $\pi(\beta|\text{data})$ is the posterior distribution of $\beta$. If the HPD interval of $\beta$ includes $0$, then we consider the predictor, corresponding to this interval, does not significantly explain the number of injuries, in the context our empirical dataset.

We organize the model fitting as follows: first, we fit the entire data with ordinary Poisson and ordinary CMP models separately. Implemented by exchange algorithm, we quickly perform the Bayesian analysis of the CMP model and it has smaller DIC than the simple Poisson model. The estimated value of $\nu$ is around $0.05<1$ indicating a large overdispersion in the data set.  Then we run the binary regression and severity regression, simultaneously, in parallel.

Table \ref{tab:simulation1} shows the estimation results from simulation 1. First, we observe that the ordinary Poisson and the ordinary CMP models have the worst performance and have difficulties recovering the true parameter values. For the binary component, the probit link which is the underlying true model gives almost identical estimates
of the regression coefficients, while the Weibull link gives the same $\beta_1$ estimate, however, with $\hat{\beta}_0$ beyond the 95\% HPD interval. The posterior mean of the shape parameter $\alpha=1.5528$. This suggests some presence of underlying dispersion in the generated data, although the probit link model has a marginally smaller DIC$=884.66$ than that of the skewed Weibull link model with DIC$=888.23$. For modeling the positive count data component for which has only 148 observations, we fit zero-truncated Poisson and zero-truncated CMP regression models. While both models give close estimates of the true regression coefficients, the ZTP provides for a marginally smaller DIC than ZTCMP. Note that the parameter estimate $\nu$ from ZTCMP model indicates that it is not statistically significantly different from 1, the case of the ordinary Poisson, which is the true underlying model in the data generation. This suggests that we can recover ordinary Poisson from CMP. Finally, considering the combination of the binary and positive components, in terms of DIC, we find that the skewed link with ZTCMP model outperforms the probit with ZTP model. This comparison suggests the promising flexibility of the skewed link and the CMP distribution models.

\begin{table}[!htbp]
\centering
\caption{Estimation results for simulation 1: true values are $\beta_0=-1$, $\beta_1=-0.5$, $\gamma_0=1$, and $\gamma_1=0.3$}
\scalebox{0.85}{
\begin{tabular}{llccccc}
          &       &       &       & \multicolumn{2}{c}{95\% HPD} &  \\ \cline{5-6} 
 \multicolumn{2}{l}{Models} &       & Estimate & Lower & Upper & DIC \\ \hline
 \multicolumn{2}{l}{Ordinary Poisson} & $\beta_0$ & -2.5259 & -2.8900 & -2.1880 & 2980.80 \\
          &       & $\beta_1$ & -0.9770 & -1.1875 & -0.7492 &  \\ \hline
 \multicolumn{2}{l}{Ordinary CMP} & $\beta_0$ & -27.1321 & -39.6127 & -16.4341 & 1861.91 \\
          &       & $\beta_1$ & -0.5069 & -5.0395 & 3.1587 &  \\ 
          &       & $\nu$ & 0.0481 & 0.0302 & 0.0697 &  \\ \hline
Binary component: & probit & $\beta_0$ & -0.9671 & -1.0667 & -0.8635 & 884.66 \\
          &        & $\beta_1$ & -0.5638 & -0.6684 & -0.4612 &  \\ [0.1cm]
          & skewed Weibull & $\beta_{0-scaled}$ & -2.0979 & -2.4435 & -1.6437 & 888.23 \\
          &        & $\beta_{1-scaled}$ & -0.6605 & -0.8771 & -0.3961 &  \\
          &        & $\alpha$ & 1.5528 & 0.9460 & 2.2530 &  \\ \hline
Positive component: & zero-truncated Poisson & $\gamma_0$ & 0.9912 & 0.8856 & 1.0977 & 647.92 \\
          &         & $\gamma_1$ & 0.2603 & 0.1556 & 0.3713 &  \\ [0.1cm]
          & zero-truncated CMP & $\gamma_0$ & 1.1332 & 1.0494 & 1.2168 & 669.53 \\
          &         & $\gamma_1$ & 0.1606 & 0.2371 & 0.4930 &  \\
          &         & $\nu$ & 1.8440 & 0.9269 & 1.5560 &  \\  \hlx{hvvv}
 \multicolumn{6}{l}{Probit+ZTP} &  1532.57 \\ \hlx{vvvhvvv}
 \multicolumn{6}{l}{Skewed Weibull+ZTCMP} &  1557.76 \\ \hlx{vvvhvvv}           
\end{tabular}}
\label{tab:simulation1}
\end{table}

Table \ref{tab:simulation2} summarizes the estimation results from simulation 2.  This simulation is generated with an overdispersion and a highly imbalanced binary outcome. According to our data generation, the true model is the hurdle model with a skewed Weibull link for the binary component and a zero-truncated CMP for the positive outcome. Our model estimation results capture this true model exceptionally well, with parameter estimates well within the true corresponding values and with the lowest DIC. As with simulation 1, both the ordinary Poisson and ordinary CMP models have the worst performance and have also difficulties recovering the parameter values. When asymmetry is introduced in the link function to introduce imbalanced binary data, the probit model does not perform well and also has difficulty with parameter recovery. Finally, for the zero-truncated Poisson, its DIC is quite competitive with a value close to the zero-truncated CMP; however, when connected with the probit model for the binary component, the probit-ZTP model does not comparatively perform as well as the skewed Weibull-ZTCMP model.

\begin{table}[!htbp]
\centering
\caption{Estimation results for simulation 2: true values are $\beta_0=-2$, $\beta_1=1$, $\alpha=3$, $\gamma_0=1$, $\gamma_1=0.3$, and $\nu=0.63$}
\scalebox{0.85}{
\begin{tabular}{llccccc}
          &       &       &       & \multicolumn{2}{c}{95\% HPD} &  \\ \cline{5-6} 
 \multicolumn{2}{l}{Models} &       & Estimate & Lower & Upper & DIC \\ \hline
 \multicolumn{2}{l}{Ordinary Poisson} & $\beta_0$ & -3.9385 & -5.6811 & -4.3548 & 3988.92 \\
          &       & $\beta_1$ & 2.8386 & 3.5905 & 4.4252 &  \\ \hline
 \multicolumn{2}{l}{Ordinary CMP} & $\beta_0$ & -19.3571 & -32.1737 & -9.0634 & 1212.34 \\
          &       & $\beta_1$ &  0.4483 & -4.8236 & 4.5379 &  \\
          &       & $\nu$ & 1.6378 & 0.5200 & 3.3325 &   \\ \hline
Binary component: & probit  & $\beta_{0-scaled}$ & -2.9467 & -3.4954 & -2.4537 & 362.02 \\
          &       & $\beta_{1-scaled}$ & 2.5592 & 2.1092 & 3.0644 &  \\[0.1cm]
          & skewed Weibull & $\beta_0$ & -1.9320 & -2.2180 & -1.5713 & 293.71 \\
          &       & $\beta_1$ & 0.9610 &  0.6799 & 1.2714 &  \\
          &       & $\alpha$ & 3.0159 & 2.0214 & 4.2118 &  \\ \hline
Positive component: & zero-truncated Poisson & $\gamma_0$ & 1.0628 & 0.8688 & 1.2915 & 755.33 \\
          &       & $\gamma_1$ & 0.2570 & 0.1288 & 0.3802 &  \\ [0.1cm]
          & zero-truncated CMP  & $\gamma_0$ & 1.0637 & 0.7940 & 1.4044 & 752.79 \\
          &       & $\gamma_1$ & 0.2130 & -0.0084 & 0.3941 &  \\
          &       & $\nu$  & 0.6497 & 0.4569 & 0.8280  &  \\ \hlx{hvvv}
 \multicolumn{6}{l}{Probit+ZTP} &  1117.35 \\ \hlx{vvvhvvv}
 \multicolumn{6}{l}{Skewed Weibull+ZTCMP} &  1046.50 \\ \hlx{vvvhvvv}           
\end{tabular}}
\label{tab:simulation2}
\end{table}

\section{Empirical application to mining injury data} \label{sec:real}
\markboth{Empirical application to mining injury data}{}

For empirical investigation of the performance of our proposed hurdle model, we analyze the data drawn from the U.S. Mine Safety and Health Administration (MSHA) for the period 2013-2016. The observations are mining companies operating in the United States and under the umbrella of the Department of Labor, MSHA has a primary mission to promote safety and health in the mining industry.

\subsection{Descriptive statistics}

\begin{figure}[!ht]
\centering
\includegraphics[width=0.48\textwidth]{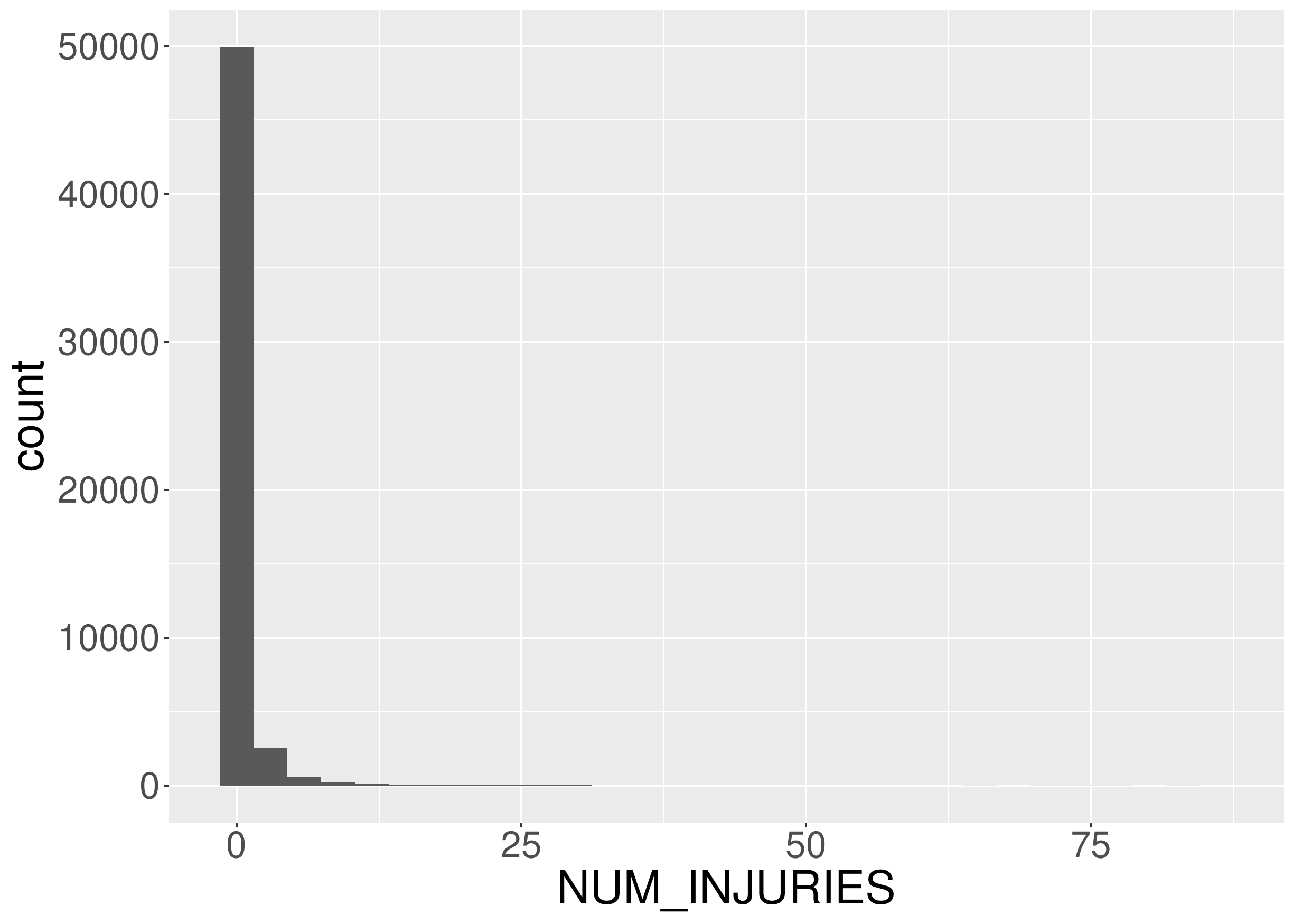}
  \hfill
\includegraphics[width=0.48\textwidth]{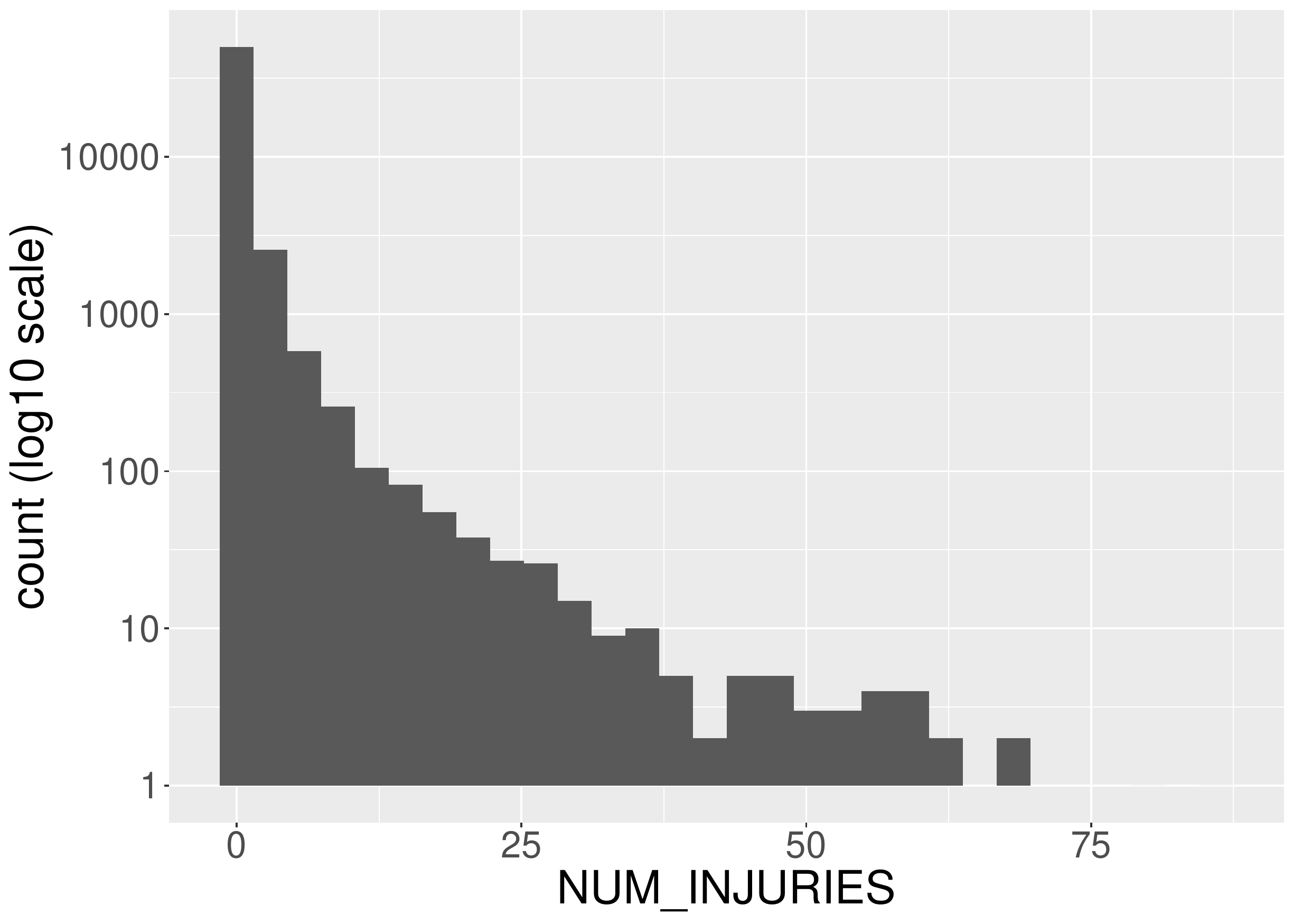}
\caption{Frequency histograms of the number of injuries on unscaled and log10 scale}
\label{fig:freq}
\end{figure}

For analysis, our main variable of interest is the number of injuries, NUM\_INJURIES, for a response variable and is considered a count variable. We have 53,746 total number of observations, with the number of positive injuries equal to 8,146; only $15\%$ on the records have occurred injuries, a highly imbalanced data. Summary statistics of NUM\_INJURIES are provided in Table \ref{tab:numinj}. To visualize the frequency of injuries observed, we present Figure \ref{fig:freq} which gives the frequency histogram of the number of injuries both on an unscaled and a log10 scale basis. The unscaled is presented to demonstrate the extremely imbalanced nature of this response variable and the log10 scale is a compression of the data to show that the positive count component cannot be trivially ignored.

\begin{table}[htbp]
\centering
\caption{Summary statistics of the number of injuries}
\scalebox{0.92}{
  \begin{tabular}{lrrrrrrrrr}
  Response variable  & MIN & 1st Q & Mean & MED & 3rd Q  & 90th & 95th & 99th & MAX \\
  \midrule
  NUM\_INJURIES & 0  &  0 & 0.4705 & 0   & 0  & 1 & 1 & 9  & 86 \\
  EMP\_HRS\_TOTAL & 1  &  1737 & 6824 & 35542   & 22180 & 62915 & 131312 & 579720   & 6811350 \\
  \bottomrule
  \multicolumn{10}{l}{* EMP\_HRS\_TOTAL was used as exposure or offset in the model.}
\end{tabular}} \label{tab:numinj}
\end{table}

For model estimation and model comparison, we partition the data into training and testing set using a random sample of 70-30 ratio, where the training  and test datasets containing 37,622 and 16,124 observations, respectively. Our dataset has a categorical attribute (MINE\_TYPE), a numerical attribute (log(SEAM)), and the 10 compositional variables consisting of percentage of employee hours spent according to the type of work done in the mining company. While some of the compositional variables have labels that may be considered self-explanatory, these variables are PCT\_HRS\_UNDERGROUND, PCT\_HRS\_SURFACE, PCT\_HRS\_STRIP, PCT\_HRS\_AUGER, PCT\_HRS\_CULM\_BANK, \, PCT\_HRS\_DREDGE, PCT\_HRS\_OTHER\_SURFACE, \, PCT\_HRS\_SHOP\_YARD, \, PCT\_HRS\_MILL\_PREP and  PCT\_HRS\_OFFICE. The values are expressed in percentages that are sparse in the sense that many components have zero percentages. As explained in Section \ref{sec:ppca}, these variables have been reconstructed using PPCA, with the optimal dimension represented by 4 components. Summary statistics are provided in Table \ref{tab:pred}. We have left out a variable describing STATES, the location of the mining company, because in our data modeling, we find this variable to be considered not statistically significant predictor of the number of injuries.

\begin{table}[!htbp]
\begin{center}
\caption{Summary statistics of the predictor variables in the mining injury dataset} \label{tab:pred}
\resizebox{\linewidth}{!}{
\begin{tabular}{llrrrrrr}
\toprule
Categorical attribute  &  Description & \multicolumn{4}{l}{}  & \multicolumn{2}{r}{Proportions} \\
\midrule
MINE\_TYPE    & \multicolumn{3}{l}{Type of mining methods.} & \multicolumn{3}{l}{Mill} & 4.71\%\\
  \multicolumn{4}{l}{}  & \multicolumn{3}{l}{Sand and Gravel} & 47.01\% \\
  \multicolumn{4}{l}{}  & \multicolumn{3}{l}{Surface} & 43.32\% \\
  \multicolumn{4}{l}{}  & \multicolumn{3}{l}{Underground} & 4.96\% \\
\midrule
Numerical attributes &  Description & Min. & 1st Q & Median & Mean & 3rd Q & Max. \\ 
\midrule
Log(SEAM) & Logarithm of the seam height. & 0 & 0 & 0 & 0.2912 & 0 & 9.2102 \\ [0.2cm]
\multicolumn{2}{l}{Principal components of the compositional variables:} &  \multicolumn{6}{l}{Percentage of employee hours} \\
\multicolumn{2}{l}{PCA1} & -1.9745 &  -0.9394 & 0.5152 & 0.0034 & 0.7638 & 0.7638 \\
\multicolumn{2}{l}{PCA2} & -2.4675 &  -0.1194 & 0.0754  & 0.0023  & 0.0779  & 2.1017   \\
\multicolumn{2}{l}{PCA3} & -2.1872 &  -0.1351 & -0.1256 & -0.0019 & 0.1673 & 1.8369 \\
\multicolumn{2}{l}{PCA4} & -2.2505 &  -0.2015 & 0.1261 & 0.0023 & 0.1565 & 2.5662 \\
\bottomrule
\end{tabular}}
\end{center}
\end{table}

\subsection{Model estimation results}

We calibrated the proposed models to the empirical dataset on mining injuries to further illustrate the flexibility of hurdle count data models with skewed link. For model comparison in order to demonstrate the robustness of these proposed models, we considered 6 different models: (a) Ordinary Poisson, (b) Ordinary CMP, (c) Probit-Zero-Truncated Poisson (Probit-ZTP), (d) Skewed Weibull-ZTP, (e) Probit-ZTCMP, and (f) Skewed Weibull-ZTCMP. All the underlying models are used to predict the count variable NUM\_INJURIES, with EMP\_HRS\_TOTAL (total employee working hours) being used as an offset, as is common when predicting a rate with variable exposure.  This means that the interpretation and prediction of the number of injuries is per employee working hours. For the categorical variable, MINE\_TYPE, the category for ``Sand and Gravel'' has been set as the reference level. Bayesian methods, as earlier discussed, are used for all models. The estimation results are tabulated in Table \ref{tab:realdata}.

\begin{table}[!htbp]
\centering
\caption{Estimation results of model fitting to the mining injury data}
\resizebox{\linewidth}{!}{
\begin{tabular}{rccc|ccc}
  \hline
 & Ordinary Poisson   & \multicolumn{2}{c|}{95\% HPD} & Ordinary CMP & \multicolumn{2}{c}{95\% HPD} \\ \cline{3-4} \cline{6-7}
 & Estimate & lower & upper & Estimate & lower & upper \\ \hline
  Intercept & -3.5527 & -3.6550 & -3.4683 & 0.3032 & 0.2456 & 0.3691 \\
  MineType\_Mill & -0.0539 & -0.2351 & 0.1277 & -0.0155 & -0.1679 & 0.1143 \\
  MineType\_Surface & 0.8523 & 0.7595 & 0.9529 & 0.2355 & 0.1486 & 0.3078 \\
  MineType\_Underground & 1.0602 & 0.7796 & 1.3614 & 0.2993 & 0.0911 & 0.5265 \\
  PCA1  & -0.7059 & -0.7536 & -0.6594 & -0.1671 & -0.2048 & -0.1289 \\
  PCA2  & 0.4331 & 0.3823 & 0.4820 & 0.0942 & 0.0560 & 0.1335 \\
  PCA3  & 0.0399 & -0.0409 & 0.1122 & 0.0719 & 0.0084 & 0.1357 \\
  PCA4  & 0.6113 & 0.5227 & 0.6916 & 0.2652 & 0.2006 & 0.3252 \\
  log(SEAM) & 0.2381 & 0.2069 & 0.2690 & 0.6862 & 0.5172 & 0.8797 \\
    $\nu$    &       &       &       & 0.6853 & 0.6502 & 0.7122 \\
    DIC   & 104886.20 &       &       & 137096.30 &       &  \\ \hline
 \multicolumn{7}{r}{}  \\ \hline
  Binary Model & Probit & \multicolumn{2}{c|}{95\% HPD} & skewed Weibull & \multicolumn{2}{c}{95\% HPD} \\ \cline{3-4} \cline{6-7}
 & Estimate* & lower & upper & Estimate & lower & upper \\ \hline
  Intercept & -0.3759 & -0.3852 & -0.3676 & -1.3530 & -1.4520 & -1.2478 \\
  MineType\_Mill & -0.0272 & -0.0534 & -0.0014 & -0.0325 & -0.0617 & -0.0003 \\
  MineType\_Surface & 0.0913 &  0.0804 & 0.1041 & 0.1149 & 0.0788 & 0.1579 \\
  MineType\_Underground &  0.1240 & 0.0812 & 0.1706 & 0.1594 & 0.0851 & 0.2344 \\
  PCA1  & -0.0922 & -0.0990 & -0.0859 & -0.1147 & -0.1458 & -0.0827 \\
  PCA2  & 0.0579 & 0.0517 &  0.0651 & 0.0728 & 0.0510 & 0.0965 \\
  PCA3  & 0.0009 & -0.0110 & 0.0103 & 0.0008 & -0.0134 & 0.0140 \\
  PCA4  & 0.0717 & 0.0609 & 0.0842 & 0.0886 & 0.0607 & 0.1140 \\
  log(SEAM) &  0.2585 & 0.2162 & 0.2974 & 0.3223 & 0.2149 & 0.4172 \\
    $\alpha$ &       &       &       & 2.9600  & 2.2704 & 3.7340 \\
    DIC   & 27679.36 &       &       &       & 27483.57 &  \\ \hline
 \multicolumn{7}{r}{}  \\ \hline
  Positive Count Model & ZTP & \multicolumn{2}{c|}{95\% HPD} & ZTCMP & \multicolumn{2}{c}{95\% HPD} \\ \cline{3-4} \cline{6-7}
 & Estimate & lower & upper & Estimate & lower & upper \\ \hline
  Intercept & -10.7023 & -10.8865 & -10.569 & -10.4312 & -10.7957 & -8.6405 \\
  MineType\_Mill & -0.7927 & -0.9933 & -0.5572 & -1.0677 & -3.3063 & -0.587 \\
  MineType\_Surface & -0.7109 & -0.8608 & -0.4948 & -0.9232 & -2.7344 & -0.5482 \\
  MineType\_Underground & -1.2351 & -1.4620 & -0.9485 & -1.5066 & -3.7361 & -0.9786 \\
  PCA1  & -0.2203 & -0.2723 & -0.1705 & -0.2423 & -0.4062 & -0.1735 \\
  PCA2  & 0.0207 & -0.0547 & 0.0883 & 0.0597 & -0.0268 & 0.3034 \\
  PCA3  & 0.1375 & 0.0152 & 0.2596 & 0.1880 & 0.0632 & 0.4883 \\
  PCA4  & 0.3305 & 0.2431 & 0.4113 & 0.3542 & 0.2746 & 0.5107 \\
  log(SEAM) & 0.0126 & -0.1522 & 0.1563 & 0.0081 & -0.0138 & 0.0573 \\
  $\nu$    &       &       &       & 1.0573 & 0.9931 & 1.2986 \\
  DIC   & 22046.87 &       &       & 21731.92 &       &  \\ \hline
        & probit+ZTP &       &       & probit +ZTCMP   &       &  \\
  DIC   & 49726.23 &       &       & 49411.20 &       &  \\
 \multicolumn{4}{r|}{} & \multicolumn{3}{c}{} \\ 
          & skewed Weibull+ZTP &       &       & skewed Weibull+ZTCMP &       &  \\
  DIC   & 49530.44 &       &       & 49215.49 &       &  \\ \hline
\multicolumn{7}{l}{* Note: Coefficients of the probit model have been scaled accordingly.} \\
\end{tabular}}
\label{tab:realdata}
\end{table}

First, when comparing the two ordinary (Poisson and CMP) models in terms of DIC, the Ordinary CMP outperforms the Ordinary CMP. According to the dispersion parameter $\nu$, there is evidence to support the presence of overdispersion. The estimated $\nu$ is 0.6853 and is well below 1 after examination of the 95\% HPD interval. For the binary component of the hurdle model, we set $\sigma=1$ for the skewed Weibull and then adjusted the scaled parameters accordingly for the probit model. In this case, we find that the skewed Weibull link model outperforms the probit model with scale. This is not at all surprising as we have seen the highly imbalanced nature of the binary outcome. We have nearly 85\% of the total observations with zero number of injuries. On the other hand, when only the positive count data is considered in the zero-truncated models, the $\nu$ parameter estimate is 1.0573 but according to the 95\% HPD interval, this estimate does not strongly support the presence of overdispersion or underdispersion. This is interesting as the overdispersion observed in the Ordinary CMP may have been primarily explained by the presence of excessive zeros. However, when the two zero-truncated models are compared, the DIC's of the probit-ZTCMP and the skewed Weibull-ZTCMP outperform the ZTP models. The DIC for the skewed Weibull is 27483.57 and that for the skewed Weibull-ZTCMP is 49215.49. The top performing model, with the lowest DIC, is the skewed Weibull link with the zero-truncated CMP model.

One primary purpose of the model construction in this paper is to understand what feature variables are important predictors of the number of injuries. Not to overwhelm the reader, we focus on the estimated coefficients of the best model, which is the skewed Weibull-ZTP model. The binary component of the hurdle model has the interpretation of the incidence of injuries, that is, an indication of the frequency of the occurrence of injuries. The positive count component has the interpretation of the severity of injuries, that is, if there is injury, the number of times the presence of injuries occurs. For the categorical variable, MINE\_TYPE, there is strong evidence to support that the type of mining methods affects both the incidence and severity of injuries. On the other hand, the continuous variable, log(SEAM), does seem to significantly affect the incidence of injuries but not the severity of the number of injuries. Finally, it is meaningless to interpret the estimated coefficients of the PCA components as their values are measured according to a weighted-average of the percentages of working hours spent on the different types of work within the mining company.

To better interpret the coefficients, we reconstruct the estimated coefficients by adjusting them back to their original scale. This can be accomplished according to the following formula:
\begin{align}\label{equ: pca}
    \text{PCA reconstruction} = \text{Estimated coefficient} \times \text{Loadings} + \text{Mean}.
\end{align}

\begin{figure}[htbp]
    \centering
    \begin{subfigure}{0.9\textwidth}
      \includegraphics[width=\linewidth]{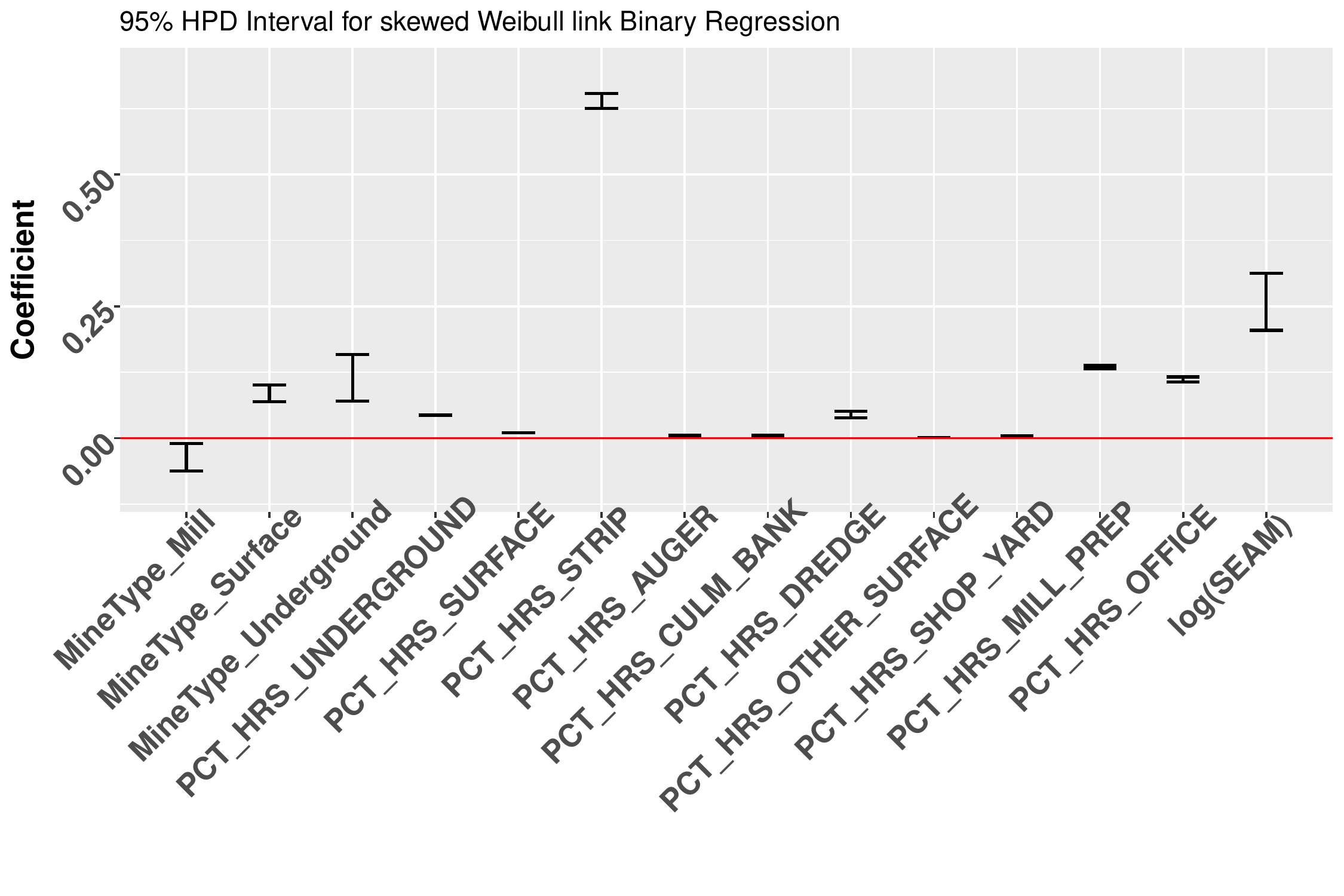}
      \label{fig:95weiCI}
    \end{subfigure}
    \begin{subfigure}{0.9\textwidth}
      \includegraphics[width=\linewidth]{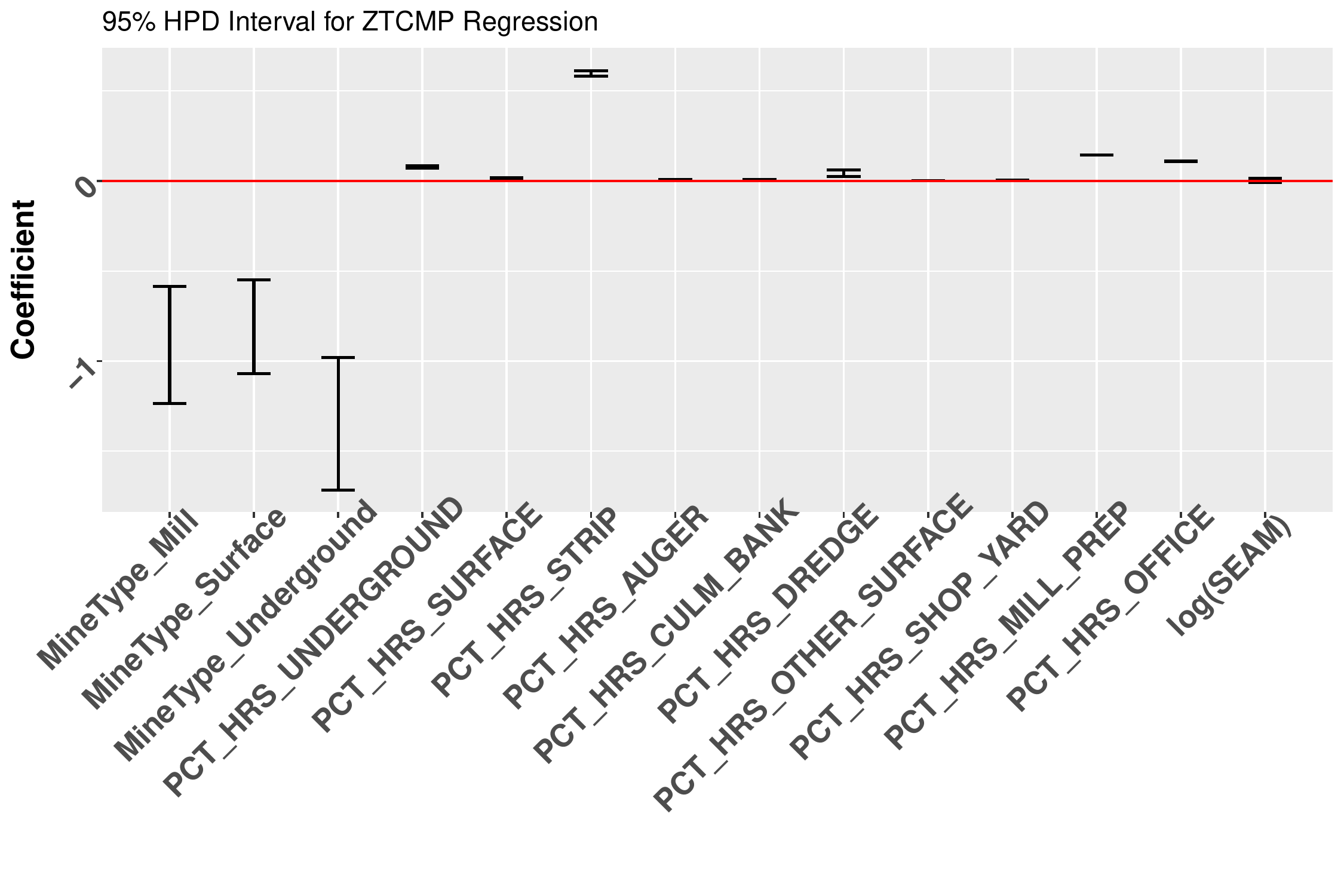}
      \label{fig:95ztpCI}
    \end{subfigure}
   \caption{95\% HPD interval for the hurdle regression model with skewed Weibull link for the zero component and the ZTCMP for the positive component}
 \label{fig:95CIbeta}
\end{figure}

Table \ref{tab:orig_pca} displays the coefficients of the original compositional data under skewed Weibull-ZTCMP regression model, calculated using equation (\ref{equ: pca}). To better visualize the effect of the predictor variables in both the incidence and severity of injuries, we present Figure \ref{fig:95CIbeta} which displays the 95\% HPD interval for the estimated coefficients. This visualization displays not only the variables that are important predictors but also provides and indication of the relative importance of the predictor variables. For example, for incidence of injuries, the PCT\_HRS\_STRIP appears to be the most important predictor, followed by log(SEAM). For severity of injuries, the MINE\_TYPE appears to be the most important predictor, followed by PCT\_HRS\_STRIP.

\begin{table}[!htbp]
\centering
\caption{Coefficient estimates of the compositional predictor variables: PCA reconstruction}
\resizebox{\linewidth}{!}{
\begin{tabular}{l||ccc||ccc}
\hline
\multirow{2}{*}{Compositional variables} & skewed & \multicolumn{2}{c||}{95\% HPD} &  zero-truncated  & \multicolumn{2}{c}{95\% HPD} \\ \cline{3-4} \cline{6-7}
  & Weibull link & lower & upper & CMP (ZTCMP) & lower & upper \\ \hline
  PCT\_HRS\_UNDERGROUND & 0.04364 & 0.04367 & 0.04439 & 0.07713 & 0.07236 & 0.08187 \\
  PCT\_HRS\_SURFACE & 0.01036 & 0.01027 & 0.01060 & 0.01578 & 0.01532 & 0.01622 \\
  PCT\_HRS\_STRIP & 0.64146 & 0.62529 & 0.65356 & 0.59558 & 0.58136 & 0.61172 \\
  PCT\_HRS\_AUGER & 0.00509 & 0.00492 & 0.00532 & 0.00630 & 0.00606 & 0.00649  \\
  PCT\_HRS\_CULM\_BANK & 0.00518 & 0.00500 & 0.00541 & 0.00638 & 0.00614 & 0.00657 \\
  PCT\_HRS\_DREDGE & 0.04475 & 0.03892 & 0.05089 & 0.04364 & 0.02418 & 0.05944 \\
  PCT\_HRS\_OTHER\_SURFACE & 0.00080 & 0.00077 & 0.00084 & 0.00096 & 0.00093 & 0.00100 \\
  PCT\_HRS\_SHOP\_YARD & 0.00391 & 0.00375 & 0.00412 & 0.00475 & 0.00456 & 0.00490 \\
  PCT\_HRS\_MILL\_PREP & 0.13361 & 0.13198 & 0.13779 & 0.14213 & 0.14223 & 0.14262 \\
  PCT\_HRS\_OFFICE & 0.11118 & 0.10644 & 0.11607 & 0.10730 & 0.10564 & 0.11029 \\ \hline
\end{tabular}}
\label{tab:orig_pca}
\end{table}

\subsection{Model validation}
The posterior predictive distribution is the distribution of possible unobserved values conditional on the observed values. For a given data set $\boldsymbol{y}=(y_1, y_2, \dots, y_n)'$, the posterior predictive distribution of a new unobserved value $y_{n+1}$ is defined to be the following
\begin{equation} \label{equ:post_predic}
f(y_{n+1}|\boldsymbol{y})  = \int f(y_{n+1}, \boldsymbol{\theta}|\boldsymbol{y}) d \boldsymbol{\theta}.
\end{equation}
If we assume the observed and unobserved data are conditional independent, given the vector of parameters $\boldsymbol{\theta}$, then formula (\ref{equ:post_predic}) can be expressed as
\begin{equation}
f(y_{n+1}|\boldsymbol{y})  = \int f(y_{n+1}|\boldsymbol{\theta}) \pi(\boldsymbol{\theta}| \boldsymbol{y}) d \boldsymbol{\theta}.
\end{equation}

MCMC samples from the posterior predictive distribution of $\boldsymbol{Y}$ can be obtained with the following procedure (\cite{hoff2009first}): For each $s \in \{1, 2, \dots, S\}$,
 \begin{itemize}
     \item draw sample from $\theta^{(s)} \sim \pi(\boldsymbol{\theta}| \boldsymbol{Y} = \boldsymbol{y}_{observed})$;
     \item draw sample from $\tilde{\boldsymbol{Y}}^{(s)} = (\tilde{y}_1^{(s)}, \dots, \tilde{y}_n^{(s)}) \sim \text{i.i.d} \ \ f(y|\theta^{(s)})$.
 \end{itemize}
Then the sequence $\{ (\theta, \tilde{\boldsymbol{Y}})^{(1)}, \dots, (\theta, \tilde{\boldsymbol{Y}})^{(S)} \}$ constitutes $S$ independent samples from the joint posterior distribution of $(\theta, \tilde{\boldsymbol{Y}})$ and the sequence $(\tilde{\boldsymbol{Y}}^{(1)}, \dots, \tilde{\boldsymbol{Y}}^{(S)})$ also constitutes $S$ independent samples generated from the posterior predictive distribution of $\tilde{\boldsymbol{Y}}$. We can use this generated samples to estimate the mean of the posterior predictive distribution and denote this estimate by $\boldsymbol{y}_{predicted}$. 

Thus to check model fit, we can generate samples from the posterior predictive distribution derived from probit-ZTP, skewed Weibull-ZTP, probit-ZTCMP, and skewed Weibull-ZTCMP models. For model validation, we utilize the unseen observations from our validation set, earlier explained how we obtained, and compare the observed to predicted values using three validation statistics to measure quality of fit:
\begin{itemize}
\item[(1)] Mean-Squared Error: $\text{MSE} = (1/m) \sum_{i=1}^m (y_{i,predicted} - y_{i,observed})^2$. Models with smaller MSE are generally preferred.
\item[(2)] Mean Absolute Error: $\text{MAE} = (1/m) \sum_{i=1}^m |y_{i,predicted} - y_{i,observed}|$. Models with smaller MAE are generally preferred.
\item[(3)] Kolmogorov-Smirnov (KS) statistic: $\text{KS} = \sup_{i} |F_{predicted}(y_i) - F_{observed}(y_i)|$. Models with smaller KS values are generally preferred.
\end{itemize}

Table (\ref{tab:prediction}) shows the competitiveness of the various models considered here. There is no single model that is considered best in terms of all three performance measures. However, although probit-ZTP outperforms all others in terms of MAE, the skewed Weibull-ZTCMP outperforms all others in terms of the MSE and the KS statistic.

\begin{table}[H]
\centering
\caption{Model performance comparison based on predicted samples}
\resizebox{\linewidth}{!}{
\begin{tabular}{c||cccc}
\hline
 Validation & \multirow{2}{*}{probit-ZTP} & \multirow{2}{*}{skewed Weibull-ZTP} & \multirow{2}{*}{probit-ZTCMP} & \multirow{2}{*}{skewed Weibull-ZTCMP} \\ 
 Measure &  &  &  &  \\ \hline
  MSE   & 4.1670   &   4.1634   &   4.1636   &  4.1600 \\
  MAE   & 0.6700  &   0.6695   &   0.6630   &  0.6701 \\
    KS statistic    & 0.0426 & 0.0426   &   0.0442 &  0.0425 \\ \hline
\end{tabular}}
\label{tab:prediction}
\end{table}

\section{Conclusion}\label{sec:conclude}
\markboth{Conclusion}{}

The models proposed in this paper have primarily been driven by the empirical data drawn from MSHA (Mine Safety and Health Administration) regarding the number of injuries incurred by mining companies. The MSHA is an agency under the umbrella of the U.S. Department of Labor that helps ensure safety in the mining industry and keeps a record of injury statistics for monitoring purposes. In our preliminary investigation of the number of injuries examined as a count data, we find peculiar characteristics of excessive zeros and the possibility of presence of overdispersion or underdispersion. We propose a class of flexible models to handle the excess zeros while simultaneously accounting for dispersion. We find that the class of hurdle count regression provides the flexibility of handling the excess zeros using a link function with additional parameter to handle the skewness. Furthermore, the Conway-Maxwell-Poisson distribution is a generalization of the ordinary Poisson with additional parameter to address dispersion. The work by \cite{song2020hotel} presents an interesting different perspective on these issues.

Furthermore, we find that the feature variables describing proportions of employee hours spent in each type of mining work (e.g., underground work, dredging, stripping, office) are considered compositional data so that the PPCA, the probabilistic counterpart of the traditional PCA, is best for processing such covariates. For inference purposes, we implemented a fully Bayesian approach to estimate the parameters in this class of hurdle CMP models. We assess the competitiveness and flexibility of this class of models against various competing models, and we find evidence of superior performance using simulation studies and the empirical application to the mining injuries. We hope that such statistical tools can help mining company management capture characteristics that contribute to mining injuries and provide guidance with respect to the notion of unfavorable factors to reduce the injury rate in real time.

\bigskip

\bibliographystyle{apalike}
\bibliography{bayesml.bib}

\begin{thebibliography}{}

\bibitem[Barcel{\'o}-Vidal et~al., 2001]{barcelo2001mathematical}
Barcel{\'o}-Vidal, C., Mart{\'\i}n-Fern{\'a}ndez, J.~A., and Pawlowsky-Glahn,
  V. (2001).
\newblock Mathematical foundations of compositional data analysis.
\newblock In {\em Proceedings of International Association of Mathematical
  Geosciences (IAMG), Cancun, Mexico}, volume~1, pages 1--20.

\bibitem[Bartholomew et~al., 2011]{bartholomew2011latent}
Bartholomew, D.~J., Knott, M., and Moustaki, I. (2011).
\newblock {\em Latent Variable Models and Factor Analysis: A Unified Approach}.
\newblock John Wiley \& Sons, West Sussex, United Kingdom, 3rd edition.

\bibitem[Cameron and Trivedi, 2013]{cameron2013}
Cameron, A.~C. and Trivedi, P.~K. (2013).
\newblock {\em Regression Analysis of Count Data}.
\newblock Cambridge University Press.

\bibitem[Caron et~al., 2018]{caron2018categorical}
Caron, R., Sinha, D., Dey, D., and Polpo, A. (2018).
\newblock Categorical data analysis using a skewed {W}eibull regression model.
\newblock {\em Entropy}, 20(3):176.

\bibitem[Chanialidis et~al., 2018]{chanialidis2018efficient}
Chanialidis, C., Evers, L., Neocleous, T., and Nobile, A. (2018).
\newblock Efficient {B}ayesian inference for {COM-P}oisson regression models.
\newblock {\em Statistics and Computing}, 28(3):595--608.

\bibitem[Chen et~al., 1999]{chen1999new}
Chen, M.-H., Dey, D.~K., and Shao, Q.-M. (1999).
\newblock A new skewed link model for dichotomous quantal response data.
\newblock {\em Journal of the American Statistical Association},
  94(448):1172--1186.

\bibitem[Chen and Shao, 1999]{chen1999monte}
Chen, M.-H. and Shao, Q.-M. (1999).
\newblock Monte {C}arlo estimation of {B}ayesian credible and hpd intervals.
\newblock {\em Journal of Computational and Graphical Statistics}, 8(1):69--92.

\bibitem[Coleman and Kerkering, 2007]{coleman2007measuring}
Coleman, P.~J. and Kerkering, J.~C. (2007).
\newblock Measuring mining safety with injury statistics: {L}ost workdays as
  indicators of risk.
\newblock {\em Journal of Safety Research}, 38(5):523--533.

\bibitem[Conway and Maxwell, 1962]{conway1962CMP}
Conway, R.~W. and Maxwell, W.~L. (1962).
\newblock A queueing model with state dependent service rate.
\newblock {\em The Journal of Industrial Engineering}, 12:132--136.

\bibitem[Dempster et~al., 1977]{dempster1977maximum}
Dempster, A.~P., Laird, N.~M., and Rubin, D.~B. (1977).
\newblock Maximum likelihood from incomplete data via the {EM} algorithm.
\newblock {\em Journal of the Royal Statistical Society: Series B
  (Methodological)}, 39(1):1--22.

\bibitem[Egozcue et~al., 2003]{egozcue2003isometric}
Egozcue, J.~J., Pawlowsky-Glahn, V., Mateu-Figueras, G., and Barcelo-Vidal, C.
  (2003).
\newblock Isometric logratio transformations for compositional data analysis.
\newblock {\em Mathematical Geology}, 35(3):279--300.

\bibitem[Fraley and Raftery, 2007]{fraley2007bayesian}
Fraley, C. and Raftery, A.~E. (2007).
\newblock Bayesian regularization for normal mixture estimation and model-based
  clustering.
\newblock {\em Journal of Classification}, 24(2):155--181.

\bibitem[Hoff, 2009]{hoff2009first}
Hoff, P.~D. (2009).
\newblock {\em A First Course in Bayesian Statistical Methods}, volume 580.
\newblock Springer, New York, New York.

\bibitem[Lambert, 1992]{lambert1992zero}
Lambert, D. (1992).
\newblock Zero-inflated {P}oisson regression, with an application to defects in
  manufacturing.
\newblock {\em Technometrics}, 34(1):1--14.

\bibitem[Li et~al., 2020]{li2020}
Li, B., Zhang, H., and He, J. (2020).
\newblock Some characterizations and properties of {COM}-{P}oisson random
  variables.
\newblock {\em Communications in Statistics - Theory and Methods},
  49(6):1311--1329.

\bibitem[Liu et~al., 2019]{liu2019risk}
Liu, Q., Meng, X., Li, X., and Luo, X. (2019).
\newblock Risk precontrol continuum and risk gradient control in underground
  coal mining.
\newblock {\em Process Safety and Environmental Protection}, 129:210--219.

\bibitem[Long, 1997]{long1997}
Long, J.~S. (1997).
\newblock {\em Regression Models for Categorical and Limited Dependent
  Variables}.
\newblock SAGE Publications, Inc.: Thousand Oaks, CA.

\bibitem[Ma and Kockelman, 2006]{ma2006bayesian}
Ma, J. and Kockelman, K.~M. (2006).
\newblock Bayesian multivariate {P}oisson regression for models of injury
  count, by severity.
\newblock {\em Transportation Research Record}, 1950(1):24--34.

\bibitem[Mc{C}ullagh and Nelder, 1989]{mccullagh1989}
Mc{C}ullagh, P. and Nelder, J.~A. (1989).
\newblock {\em Generalized Linear Models}.
\newblock Chapman and Hall: London.

\bibitem[McLachlan and Krishnan, 2008]{mclachlan2008}
McLachlan, G.~J. and Krishnan, T. (2008).
\newblock {\em The EM Algorithm and Extensions}.
\newblock John Wiley \& Sons, Hoboken, New Jersey.

\bibitem[Minka et~al., 2003]{minka2003computing}
Minka, T.~P., Shmueli, G., Kadane, J.~B., Borle, S., and Boatwright, P. (2003).
\newblock Computing with the {COM}-{P}oisson distribution.
\newblock Research Showcase @ CMU, Department of Statistics and Dietrich
  College of Humanities and Social Sciences, Carnegie Mellon University.

\bibitem[Mullahy, 1986]{mullahy1986hurdle}
Mullahy, J. (1986).
\newblock Specification and testing of some modified count data models.
\newblock {\em Journal of Econometrics}, 33:341--365.

\bibitem[Nowrouzi et~al., 2017]{nowrouzi2017bibliometric}
Nowrouzi, B., Rojkova, M., Casole, J., and Nowrouzi-Kia, B. (2017).
\newblock A bibliometric review of the most cited literature related to mining
  injuries.
\newblock {\em International Journal of Mining, Reclamation and Environment},
  31(4):276--285.

\bibitem[Nyamundanda et~al., 2010]{nyamundanda2010probabilistic}
Nyamundanda, G., Brennan, L., and Gormley, I.~C. (2010).
\newblock Probabilistic principal component analysis for metabolomic data.
\newblock {\em BMC Bioinformatics}, 11(1):571.

\bibitem[Paul, 2009]{paul2009predictors}
Paul, P.~S. (2009).
\newblock Predictors of work injury in underground mines—an application of a
  logistic regression model.
\newblock {\em Mining Science and Technology (China)}, 19(3):282--289.

\bibitem[Sarul and Shin, 2015]{sarul2015genins}
Sarul, L.~S. and Shin, S. (2015).
\newblock An application of claim frequency data using zero inflated and hurdle
  models in general insurance.
\newblock {\em Journal of Business, Economics and Finance}, 4(4):732--743.

\bibitem[Shmueli et~al., 2005]{shmueli2005useful}
Shmueli, G., Minka, T.~P., Kadane, J.~B., Borle, S., and Boatwright, P. (2005).
\newblock A useful distribution for fitting discrete data: revival of the
  {C}onway-{M}axwell-{P}oisson distribution.
\newblock {\em Journal of the Royal Statistical Society: Series C (Applied
  Statistics)}, 54(1):127--142.

\bibitem[Shmueli et~al., 2004]{shmueli2004modeling}
Shmueli, G., Russo, R.~P., and Jank, W. (2004).
\newblock Modeling bid arrivals in online auctions.
\newblock {\em Robert H. Smith School Research Paper No. RHS-06-001}.

\bibitem[Song, 2020]{song2020hotel}
Song, K.-S. (2020).
\newblock Simultaneous statistical modelling of excess zero,
  over/underdispersion, and multimodality with applications in hotel industry.
\newblock {\em Journal of Applied Statistics}.
\newblock in press, DOI: 10.1080/02664763.2020.1769577.

\bibitem[Spiegelhalter et~al., 2002]{spiegelhalter2002bayesian}
Spiegelhalter, D.~J., Best, N.~G., Carlin, B.~P., and Van Der~Linde, A. (2002).
\newblock Bayesian measures of model complexity and fit.
\newblock {\em Journal of the Royal Statistical Society: Series B (Statistical
  Methodology)}, 64(4):583--639.

\bibitem[Tipping and Bishop, 1999]{tipping1999probabilistic}
Tipping, M.~E. and Bishop, C.~M. (1999).
\newblock Probabilistic principal component analysis.
\newblock {\em Journal of the Royal Statistical Society: Series B (Statistical
  Methodology)}, 61(3):611--622.

\bibitem[Yin et~al., 2020]{Yin2020skewed}
Yin, S., Dey, D.~K., Valdez, E.~A., Gan, G., and Vadiveloo, J. (2020).
\newblock Skewed link regression models for imbalanced binary response with
  applications to life insurance.
\newblock arXiv:2007.15172.

\bibitem[Zorn, 1998]{zorn1998socio}
Zorn, C.~J. (1998).
\newblock An analytic and empirical examination of zero-inflated and hurdle
  {P}oisson specifications.
\newblock {\em Sociological Methods \& Research}, 8(3):368--400.

\end{thebibliography}

\end{document}